\documentclass[hyperref]{article}
\usepackage{makeidx}
\usepackage[colorlinks=true,citecolor=black]{hyperref}
\usepackage{hyperref}
\usepackage{multirow}
\usepackage{multicol}
\usepackage[dvipsnames,svgnames,table]{xcolor}
\usepackage{graphicx}
\usepackage{epstopdf}
\usepackage{ulem}
\usepackage{hyperref}
\usepackage{amsmath}
\usepackage{amssymb}
\usepackage{enumerate}
\usepackage{tabulary}
\usepackage[a4paper]{geometry}
\usepackage{subfigure}
\usepackage{caption}
\usepackage{tabularx}
\usepackage{makecell}
\usepackage[justification=centering]{caption}
\makeatother
\hypersetup{
	colorlinks=true,
	linkcolor=black
}
\begin{document}
	\captionsetup[figure]{labelfont={bf},name={Fig.},labelsep=period}
	\title{\bf\Large Experimental Study on CTL model checking using Machine Learning}
	
	\date{}
	\author{\large Weijun ZHU\\
		{\sffamily\small School of Information Engineering, Zhengzhou University, Zhengzhou City,450001, China}\\
	}

	\maketitle
	{\noindent\small{\bf Abstract:}
		The existing core methods, which are employed by the popular CTL model checking tools, are facing the famous state explode problem. In our previous study, a method based on the Machine Learning (ML) algorithms was proposed to address this problem. However, the accuracy is not satisfactory. First, we conduct a comprehensive experiment on Graph Lab to seek the optimal accuracy using the five machine learning algorithms. Second, given the optimal accuracy, the average time is seeked. The results show that the Logistic Regressive (LR)-based approach can simulate CTL model checking with the accuracy of 98.8\%, and its average efficiency is 459 times higher than that of the existing method, as well as the Boosted Tree (BT)-based approach can simulate CTL model checking with the accuracy of 98.7\%, and its average efficiency is 639 times higher than that of the existing method. }
	
	\vspace{1ex}
	
	\section{Introduction}
	Model checking was presented by Turing Award winner Prof. Clarke et al \cite{1}. It has been widely used in several fields such as CPU design, network protocols. The basic principle of model checking can be depicted as follows: (1) a Kripke structure is employed to construct a systematic model, while a formula of a temporal logic is employed to describe a property which should be satisfied by this system; (2) a model checking algorithm decides whether the Kripke structure satisfies the formula or no. In model checking, linear temporal logic \cite{2} and computational tree logic (CTL) \cite{3} are the two popular temporal logics. In other words, CTL model checking and LTL model checking are the two different problems which are related with each other.
	
	The state explosion problem is always one of the important bottlenecks of CTL model checking. Unfortunately, the state explosion problem inherently originates from the gene of model checking. Thus, we proposed an approximate CTL model checking approach using machine learning in our previous study \cite{4}\cite{5}. And we obtained the accuracy of 96\% when we predict the result of CTL model checking via ML algorithms. Considering model checking is a formal method which is very accurate, the accuracy of 96\% is not enough to formal computing. How to further improve the accuracy of ML-based approximate CTL model checking? Something more needs to be done. This is the motivation of this paper.	
	\section{The principle of the approximate CTL model checking based on machine learning}
	Given a Kripke structure and a CTL formula, how to decide whether this Kripke structure satisfy this formula or not? This is a model checking problem, as well as a ML binary classification problem. If a model checking algorithm is used, the state explode will be inevitable. If a ML algorithm is employed, the state explode will not happen.

     The principle of the ML-based approximate CTL model checking is as follows. (1) A number of records in a data set are inputted to a ML algorithm for training, where each record has the two features, one is Kripke structure and the other is CTL formula, as well as each record has only one label, i.e., the result of CTL model checking (yes or no). (2) If a record in a test set is inputted to the trained ML model M, the predictive result will be gotten. 

     See \cite{4}\cite{5} for more details on the ML-based approximate CTL model checking.	
	\section{Simulated experiments}
	\subsection{Experimental objective}
	We will compare the ability and the efficiency between the ML-based approximate CTL model checking and the classic CTL model checking.
	\subsection{The simulation platform}
	\begin{enumerate}[(1)]
		\item CPU: Intel(R) Core(TM) i7-4790 CPU @3.60GHz.
		\item RAM: 8.0G RAM.
		\item OS: Windows 10.
		\item NuSMV \cite{6}: for performing CTL model checking.
		\item Graph Lab \cite{7}: for implementing Random Forest (RF), Boosted Tree (BT), K-Nearest Neighbor (KNN), Decision Tree (DT) and Logistic Regressive (LR) algorithms.
          \item the raw data: same as the one in Ref.\cite{4}\cite{5}.
	\end{enumerate}
	\subsection{Experimental Procedures}
	Similar with that of Ref.\cite{4}\cite{5}.
		
	\begin{figure}
		\centering
		\subfigure[result of prediction via RF ]{
			\begin{minipage}[b]{0.35\textwidth}
				\includegraphics[width=\textwidth]{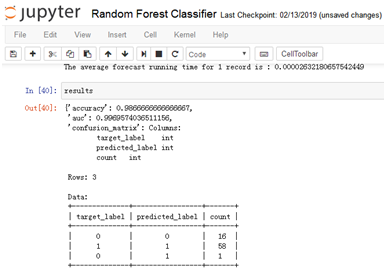} \\
				\label{fig1-1}
			\end{minipage}
		}
	
		\subfigure[ result of prediction via BT ]
		{
			\begin{minipage}[b]{0.35\textwidth}
				\includegraphics[width=\textwidth]{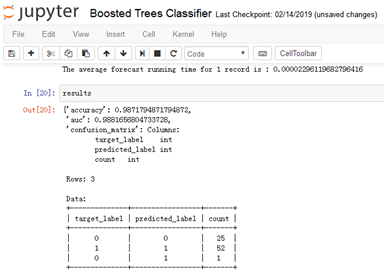} \\
				\label{fig1-2}
			\end{minipage}
		}

		\subfigure[result of prediction via KNN ]{
			\begin{minipage}[b]{0.35\textwidth}
				\includegraphics[width=\textwidth]{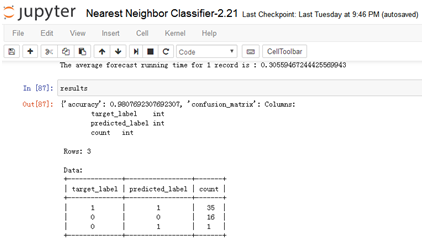} \\
				\label{fig1-3}
			\end{minipage}
		}

		\subfigure[result of prediction via DT ]{
			\begin{minipage}[b]{0.35\textwidth}
				\includegraphics[width=\textwidth]{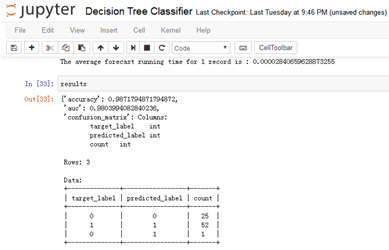} \\
				\label{fig1-4}
			\end{minipage}
		}

		\subfigure[result of prediction via LR]{
			\begin{minipage}[b]{0.35\textwidth}
				\includegraphics[width=\textwidth]{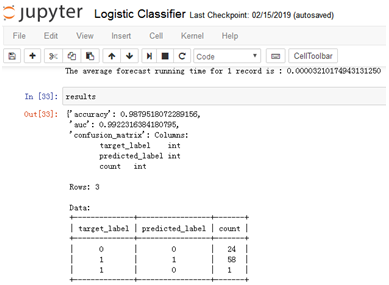} \\
				\label{fig1-5}
			\end{minipage}
		}
		\caption{Result of prediction based on different ML algorithms, there are 400 records in sample set}
		\label{fig1}
	\end{figure}
	
	\subsection{Experimental Results}
The same raw data with the one in Ref.\cite{4}\cite{5} are inputted to Graphlab. Fig.1 illustrates the result of ML prediction including predictive accuracy and average time for predicting the model checking result for a pair of Kripke structure and a CTL formula. And Tab.1, Tab.2, Tab.3, Tab.4 and Tab.5 show the value of parameters when the above predictive results occur. Furthermore, Tab.6 gives a comparison between the ML-based approaches and a typical CTL model checking approach, i.e., NuSMV. 

As shown in these figures and tables, the accuracy of the LR-based approach is 98.8\%, and the accuracy of the BT-based approach is 98.7\%. It indicates that predictions are very accurate using the ML-based method to simulate CTL model checking.
	\begin{table}
		\centering		
		 {\caption{  Graphlab experiment where length of each CTL formula is 500 \protect \\ What are the values of the parameters if the illustrations of Fig.1(a) occur}}
		
		\begin{tabular}{ccc}
			\hline
			Names of parameters & Meaning of parameters  & Values of parameters \\
			\hline
			seed  & \makecell{Seed for the random number generator used to split} & 459 \\
			fraction  & \makecell{For determining the proportion of the records of training set
				\\ in the total records of data set}
			 & 0.8 \\
			\hline
		\end{tabular}
	\end{table}

	\begin{table}
		\centering		
		 {\caption{  Graphlab experiment where length of each CTL formula is 500 \protect \\ What are the values of the parameters if the illustrations of Fig.1(b) occur}}
		
		\begin{tabular}{ccc}
			\hline
			Names of parameters & Meaning of parameters  & Values of parameters \\
			\hline
			seed  & \makecell{Seed for the random number generator used to split} & 536 \\
			fraction  & \makecell{For determining the proportion of the records of training set
				\\ in the total records of data set}
			 & 0.8 \\
			\hline
		\end{tabular}
	\end{table}

	\begin{table}
		\centering		
		 {\caption{  Graphlab experiment where length of each CTL formula is 500 \protect \\ What are the values of the parameters if the illustrations of Fig.1(c) occur}}
		
		\begin{tabular}{ccc}
			\hline
			Names of parameters & Meaning of parameters  & Values of parameters \\
			\hline
			seed  & \makecell{Seed for the random number generator used to split} & 399 \\
			fraction  & \makecell{For determining the proportion of the records of training set
				\\ in the total records of data set}
			 & 0.86 \\
			\hline
		\end{tabular}
	\end{table}

	\begin{table}
		\centering		
		 {\caption{  Graphlab experiment where length of each CTL formula is 500 \protect \\ What are the values of the parameters if the illustrations of Fig.1(d) occur}}
		
		\begin{tabular}{ccc}
			\hline
			Names of parameters & Meaning of parameters  & Values of parameters \\
			\hline
			seed  & \makecell{Seed for the random number generator used to split} & 536 \\
			fraction  & \makecell{For determining the proportion of the records of training set
				\\ in the total records of data set}
			 & 0.8 \\
			\hline
		\end{tabular}
	\end{table}

	\begin{table}
		\centering		
		 {\caption{  Graphlab experiment where length of each CTL formula is 500 \protect \\ What are the values of the parameters if the illustrations of Fig.1(e) occur}}
		
		\begin{tabular}{ccc}
			\hline
			Names of parameters & Meaning of parameters  & Values of parameters \\
			\hline
			seed  & \makecell{Seed for the random number generator used to split} & 2077 \\
			fraction  & \makecell{For determining the proportion of the records of training set
				\\ in the total records of data set}
			 & 0.8 \\
			\hline
		\end{tabular}
	\end{table}

	\begin{table}
		\centering		
		 {\caption{ The new method improves the efficiency}}
		
\begin{tabular}{cp{4cm}<{\centering}p{4cm}<{\centering}cc}
\hline
ML algorithms & Average running time $t_1$ (in Seconds) of NuSMV for one pair of Kripke structure and formula & Average predictive time $t_2$ (in Seconds) of ML for one record & $t_1$/$t_2$ \\
\hline
RF & 0.0147 & 0.000026 & 565 \\
BT & 0.0147 & 0.000023 & 639 \\
KNN & 0.0147 & 0.3056 & 0.05 \\
DT & 0.0147 & 0.000028 & 525 \\
LR & 0.0147 & 0.000032 & 459 \\
\hline
\end{tabular}
	\end{table}
	
	\section{Conclusions}
	In this study, we obtain a comprehensive experimental result on the ML-based approximate CTL model checking. This method has a high accuracy of approximate CTL model checking which has declined slightly (1.2\%-1.3\% only) from that of actual CTL model checking, in exchange for a substantial increase in efficiency of CTL model checking (639 times).
	\section*{Acknowledgements}
	This work has been supported by the National Natural Science Foundation of China (No.U1204608).

\end{document}